\documentclass[rnote]{aa} 
\usepackage{graphicx}
\usepackage{txfonts}
\usepackage[figuresright]{rotating}
\begin{document}
   \title{Interferometric observations of the multiple stellar system $\delta$\,Velorum}

   \author{A. Kellerer
          \inst{1}
          \and
          M.\ G.\ Petr-Gotzens\inst{1}
          \and
          P. Kervella\inst{2}
          \and
          V. Coud\'e du Foresto\inst{2}
          }
   \offprints{A. Kellerer}

   \institute{European Southern Observatory, Karl-Schwarzschild-Str.\ 2,
              D-85748 Garching, Germany\\
              \email{aglae.kellerer@eso.org, mpetr@eso.org}
          \and
            LESIA, Paris observatory\\
            \email{vincent.foresto@obspm.fr, pierre.kervella@obspm.fr}   
             }

   \date{Received --; accepted ---}

 \abstract
   {$\delta$\,Velorum is a nearby ($\sim$24\,pc) triple stellar system,
containing a close, eclipsing binary (Aa, Ab) discovered in 2000.  }
   {Multiple systems provide an opportunity to determine the set of fundamental parameters (mass, luminosity, size and chemical composition) of  coeval stars.
   These parameters can be obtained with particular precision in the case of eclipsing binaries; 
   for $\delta$\,Velorum's components (Aa, Ab) this potential has however not yet been exploited. }
   {We have analyzed interferometric observations of the close binary (Aa, Ab),  obtained with the VINCI instrument and two VLTI siderostats. The measurements, which resolve the two components for the first time, are fitted onto the simple model of two uniformly bright, spherical stars.}
   {The observations suggest that Aa and Ab have larger diameters than expected if they were on the main sequence, and that they are, thus, in a later evolutionary state. }
   {}


   \maketitle
%

\section{Introduction}

$\delta$\,Velorum (HD\,74956) is one of the fifty
brightest stars on the sky, with a visual magnitude of $m_{\rm V\/}=1.96$ mag
(Johnson et al.\ \cite{Johnson}).
It is a multiple stellar system 
(e.g.\ Worley \& Douglass \cite{worley97}), but
in spite of its brightness and near distance, $\pi=(40.90\pm 0.38)$\,mas
(Perryman et al.\ \cite{ESA}), the issue of its 
composition remains unresolved. As early as 1847,  Herschel published his  detection of
two faint visual companions, $\delta$~Vel~C and D, at distance $69^{\prime\prime}$ apart 
from $\delta$~Vel~A.
Another companion, $\delta$~Vel~B -- at the time separated by $\sim3\arcsec$ from 
$\delta$~Vel~A -- was later discovered by Innes (\cite{Innes1895}). 
The separation $0\farcs736\pm0\farcs014$ between the components A and B
appeared surprising when measured by {\it Hipparcos},
but it was well explained later, in terms of the orbit computation of Argyle et al.
(\cite{Argyle2002}) that showed a highly  elliptical orbit of component B with period 
$P=142\,$yr.
In 1979, preliminary results from speckle interferometry suggested an even 
further component of the system (Tango et al.\ \cite{tango}). This apparent companion  
was found at a separation of $\sim 0\farcs6$ and was taken to be a further component, 
because for star B the separation at the time was believed to be $\sim 3^{\prime\prime}$.
By now, however, it seems very likely that the speckle observations resolved 
$\delta$~Vel~B; while there is still an unexplained disagreement for the position angle, the measured small separation does fit well with the orbital solution found by Argyle et al.
(\cite{Argyle2002}).
As noted in earlier publications, (Hoffleit et al. \cite{Hoffleit}, Otero at al. \cite{Otero}), the two
stars that are currently termed $\delta$~Vel~C and D were taken to be associated with the pair AB because of seemingly similar proper motion. However, we have not found the source of the proper motion measurement of C and D.  Finally, the most luminous component, A, was recently recognized to be a close eclipsing binary with a period $T=45.15$\,days
(Otero et al. \cite{Otero}). 
$\delta$\,Vel has since been classified as a quintuple stellar system. 

This investigation is focussed on the bright eclipsing binary, $\delta$\,Vel\,A, but we also argue that $\delta$\,Vel\,C and D are not physically associated with $\delta$\,Vel\,A,B. While this makes $\delta$\,Vel a  triple system, it takes away little from its challenging potential to obtain important information on stellar evolution. As the inclination, $i$, of its orbital plane is constrained to be close to $90^{\circ}$, an eclipsing binary system provides one of the best means to obtain, in terms of the Kepler laws of motion, fundamental stellar parameters.

In this research note we present the first interferometric observations of the eclipsing binary $\delta$\,Vel\,A, obtained with ESO's Very Large Telescope Interferometer (VLTI) and its ``commissioning instrument" VINCI. The measurements resolve, for the first time, this binary system. They are here analyzed with non-linear least-square fitting methods. We combine our interferometric results with existing photometric and spectroscopic observations, estimate some orbital parameters of the $\delta$\,Vel\,A binary system, and discuss, based on the results, the stellar properties of the individual components.


\section{Characteristics of $\delta$~Vel~A derived from previous measurements}\label{Photometry}

\subsection{Orbit orientation and eccentricity}
\label{Orbitalelements}

\begin{figure}
   \includegraphics[angle=90,width=8cm]{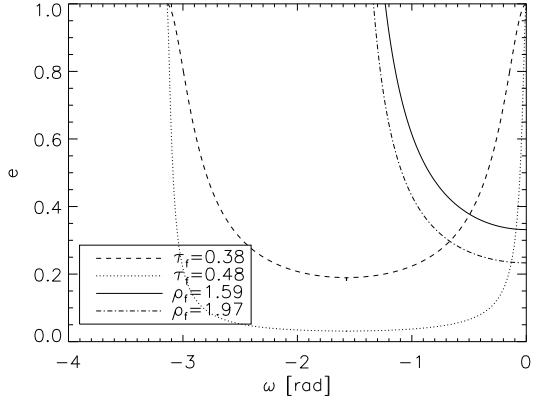}
      \caption{The angle, $\omega$, at primary eclipse and the eccentricity, $e$, are constrained 
      by the fractional durations between the eclipses, $\tau_f=0.43\pm0.05$, and the fractional durations of the eclipses, $\rho_f=1.78\pm0.19$. }
\label{omega2exc}
   \end{figure}

In this subsection, {\it a priori} estimates of two orbital parameters of 
the $\delta$\,Vel\,(Aa-Ab) system are derived
from the time interval between the eclipses and their durations,
the eccentricity, $e$, and the angle, $\omega$, between the semi-major axis
and the line of sight. The angle $\omega$ is similar to,
but must not 
be confused with the more generally used parameter {\it longitude of 
periastron.}

As reported by Otero et al.\,(\cite{Otero}) and Otero (\cite{Otero2006}), 
the fractional orbital period from the primary to the secondary eclipse equals  
$\tau_f=0.43\pm0.05$.
The secondary eclipse has been observed by the Galileo satellite in 1989,
and its duration and depth have been fairly precisely established to be $0.91\pm0.01$\,days  
and $\Delta m_{II}=0.32\pm0.02$ (Otero\,\cite{Otero2006}).
The same spacecraft observed the primary eclipse several years later, although
its measurements had become less accurate then.
The approximate duration and depth of the primary eclipse are 
$0.51\pm0.05$\,days  and $\Delta m_{I}=0.51\pm0.05$
(Otero\,\cite{Otero2006}).
The ratio of durations thus amounts to $\rho_f=1.78\pm0.19$.

The relative motion of the two stars $\delta$\,Vel Aa and Ab is taken to be 
independent of external forces, and the vector, {\bf s\/}, from Ab to Aa 
traces an elliptical orbit around Ab as a focal point.
As the photometric light curve indicates a total eclipse for 
$\delta$\,Vel~A,
the inclination of the orbit needs to be close to 90$^\circ$. 
To simplify the equations, we assume $i=90^\circ$.
Ab is taken to be the star with the higher surface brightness. 
During the primary eclipse, which is deeper,  Ab is thus eclipsed by Aa.
The angle $\theta$ of {\bf s\/}, also called the {\it true anomaly\/}, is 
zero at periastron and increases to $\pi$ as the star
 moves towards apastron.
The distance between the stars depends on $\theta$ according to the relation:
\begin{equation}
s(\theta) = a(1 - e^{2})/(1+e \cos(\theta))
\end{equation}
where $a$ denotes the semi-major axis.
In line with Kepler's second law the vector {\bf s\/} covers equal areas per unit time.
The fractional orbital period to reach angle $\theta$ is, accordingly:
\begin{eqnarray}
\tau (\theta)&=& \frac{2}{A} \int_{0}^{\theta} s^{2}(\theta') {\rm d\/}\theta'\\
&=&[2 {\rm arctan\/}(f_{1} \tan(\theta/2) - f_{2} \sin(\theta)/(1+
e \cos(\theta))] / 2\pi
\end{eqnarray}
where $A = \pi a^{2} \sqrt{(1 - e^{2})}$ equals the area of the ellipse.
$f_{1} = \sqrt{((1-e)/(1+e))}$ and
$f_{2} = e \sqrt{(1-e^{2})}$.

During the primary eclipse, when  
the star with the lower surface brightness, Aa, covers Ab, the vector {\bf s\/} is directed 
towards Earth, 
and $\theta$ equals
$\omega$. During the secondary eclipse $\theta$ equals $\omega+\pi$.
Thus:
\begin{eqnarray}
\tau_f &=& 0.43 \pm0.05\\
&=& \int_{\omega}^{\omega+\pi} {\rm d\/}\theta/(1+e \cos(\theta))^{2}  \\
&=& [{\rm arctan\/}(f_{1} \tan((\omega+\pi)/2))-{\rm arctan\/}(f_{1}
 \tan (\omega/2)) \nonumber\\
&+&f_{2} \sin(\omega)/(1-e^{2} \cos(\omega)^{2})] / \pi
\label{Eq:tau}
\end{eqnarray}
 which determines $\omega$ for any given eccentricity, $e$ (Fig.\,\ref{omega2exc}).
The orbital velocity decreases as $\theta$ goes from zero to $\pi$, 
i.e. from the periastron to the apastron. In the subsequent interval, $\pi$ to 2$\pi$ 
(or -$\pi$ to 0), it increases again.
If the line of sight contained the orbital major axis, i.e. $\omega= 0$ or $\pi$,
the fractional duration between eclipses $\tau$  would equal 0.5.
Note that for such values of $\omega$, Eq.\,\ref{Eq:tau} is not defined,
yet $\tau$ tends towards 0.5 when $\omega$ approaches $0$, or $\pi$.
If the line of sight contained the orbital minor axis, i.e. $\omega= \pi$/2 or -$\pi$/2,
the maximum and minimum values of $\tau$ would be reached.
Values of $\tau$ less than 0.5 are thus associated with negative 
$\omega$ values. 
Since the fractional orbital period from the primary to the secondary eclipse
is 0.43, the angle $\omega$ must lie between -$\pi$ and 0.
As Fig.\,\ref{omega2exc} shows, the eccentricity needs to be 
larger than $\approx 0.03$.

On the other hand, $\omega$ can be further constrained through the ratio
of the eclipse durations as follows. 
The eclipse durations are inversely proportional to the product 
$r\,{\rm d\/}\theta/{\rm d\/}t$
of radius and angular velocities during the eclipses. 
They are thus proportional to $s(\theta$), and their ratio is:
 \begin{eqnarray}
 \rho(\omega) &=& (1 - e \cos(\omega)) / (1 + e \cos(\omega))
 \end{eqnarray}
Given $\rho_f = 1.78\pm0.19$, this leads to a second relation between $e$ and 
$\omega$. As illustrated by Fig.\,\ref{omega2exc}, simultaneous agreement with 
both observed 
values $\tau_f$  and $\rho_f$
is reached only if $e\;\in\;[0.23 - 0.37]$ and $\omega\;\in\;-[0.1 - 0.7]$\,rad.

\subsection{Semi-major axis and stellar parameters}
\label{Stellarparameters}

Orbital motion in the triple system $\delta$\,Vel(Aa+Ab+B) has  recently been substantiated and analyzed by  Argyle et al.\ (\cite{Argyle2002}). From position measurements
taken over a period of roughly 100\,years the authors inferred a 
${\rm P} = 142$\,yr orbit for component B, and deduced a total dynamical mass 
$M(Aa)+M(Ab)+M(B)=5.7^{+1.27}_{-1.08}\;M_{\odot}$. 
Photometric and spectroscopic measurements
of the individual components being few and partly inconclusive, 
individual mass estimates are still difficult. 

{\it Hipparcos} 
measured an apparent magnitude of $H_{\rm P}=1.991$ for $\delta$\,Vel\,A
and $H_{\rm P}=5.570$ for $\delta$\,Vel\,B. With the transformations given
by Harmanec (\cite{Harmanec}), the approximate Johnson V magnitudes are
$m_{\rm V\/} = 1.99$ and $m_{\rm V\/} = 5.5$ for $\delta$\,Vel\,A and
$\delta$\,Vel\,B, respectively. But  the colours of the 
individual 
$\delta$\,Vel components being unknown, it needs to be noted that the uncertainty of $m_{\rm V\/}$
can be as high as $\sim0.07$mag.  
Since $\delta$\,Vel\, is close ($d = 24.45$\,pc according to {\it Hipparcos})
no interstellar reddening towards the source needs to be assumed, and hence, 
the absolute magnitudes are
$M_{\rm V\/} \sim 0.05$ for $\delta$\,Vel\,A and $M_{\rm V\/} \sim 3.6$ for
$\delta$\,Vel\,B.

Several authors have analyzed spectra of $\delta$\,Vel\,A 
(e.g. Wright \cite{Wright}, Alekseeva \cite{Alekseeva}, Levato \cite{Levato},  
Gray \& Garrison \cite{GrayGarrison}). Many of their measurements have probably included
 $\delta$\,Vel\,B, but its flux is too low to add a 
significant contribution.
From the metal line ratios and Balmer line equivalent widths all
authors deduced either spectral type A0\,V or A1\,V.
This being most likely an average classification of the two stars, Aa and Ab,
one star should be slightly hotter and the other cooler than a A0/1V star.
No signatures of a double-lined spectroscopic binary were reported in 
any of the spectroscopic observations. 

Based on the spectrophotometric 
information referred to above, and under the assumption
that all $\delta$\,Vel components are on the main sequence, it is
suggested that Aa and Ab are of spectral type between A0V and A5V with 
masses in the range 2.0--3.0$M_{\odot}$. Furthermore it follows
 that B is an F-dwarf with mass about 
$\sim 1.5M_{\odot}$. This agrees reasonably with the total dynamical
mass derived by Argyle et al.\ (\cite{Argyle2002}).

An {\it a priori} estimate of the semi-major axis, $a$, of the Aa-Ab system is 
next derived from the mass sum of Aa+Ab
($5\pm1\,M_\odot$) 
and its orbital period ($T=45.150\pm0.001$\,days),
which leads to $a = (6.4\pm0.5)\times 10^{10}$\,m=\,$0.43\pm0.04$\,AU.\\
If they are main sequence early A stars,
Aa and Ab should have  stellar diameters between $1.7-2.4\,D_{\odot}$.

Finally, the depths of the eclipses can be used to constrain the 
surface brightness ratio $\phi$ of the two eclipsing components,
$\delta$\,Vel Aa and Ab,
\begin{equation}\label{eq:phi}
1.28 \leq \phi = \frac{1-10^{-\Delta m_{I}/2.5}}{1-10^{-\Delta m_{II}/2.5}}\leq 1.67
\end{equation}


\section{VLT Interferometer/VINCI Observations}
\subsection{Data description}\label{Datadescription}
During April-May 2003, the ESO Very Large Telescope Interferometer (VLTI) was used to observe the eclipsing binary $\delta$\,Vel(Aa+Ab)
in the K-band at four orbital phases with the single-mode fiber 
based instrument VINCI (Glindemann\,\cite{Glindemann}, 
Kervella et al.\,\cite{Kervella2003}). 
The observations were performed  with two siderostats, placed at 
stations B3 and M0, separated by 155.368 m.
Table\,\ref{Observational_data} summarizes the resulting data.\\
Every interferometric observation yields a fringe contrast or 
squared visibility, $V^2$,
whose variations are not only due to interferometric modulation, but also
to atmospheric and instrumental fluctuations. Accordingly
the raw squared visibilities need to be calibrated by a reference star. 
To this purpose the observations of $\delta$\,Vel were combined with observations of HD\,63744,
a star of spectral type K0III, with an estimated diameter of 
$1.63\pm0.03$\,mas (Bord\'e et al.\,\cite{Borde}). 
The interferometric measurements were then analyzed by use of the VINCI data 
reduction pipeline, described in detail in Kervella et
al.\,(\cite{Kervella2003}).

\begin{table}
\caption{Details of the VINCI measurements: observing date, orbital phase of
$\delta$\,Vel\,(Aa+Ab), calibrated squared visibility $V^2$ and standard deviation
$\sigma_{V^2}$, number of accepted scans $N_S$ (out of 500).
The uncertainty on the phase determination equals $\pm0.002$.}
\label{Observational_data}
\flushright
\begin{tabular}{l rl rl rlr rlr rl}
\hline\hline
Date & Julian Date - 2452700 &Phase&$V^2\pm\sigma_{V^2}, \%$&$N_s$\\ \hline
21 Apr 03 & 50.628 &0.937 &57.40$\pm$3.60&383\\
 &50.633&0.937 &54.20$\pm$3.60&298\\
 & 50.639 &0.937 &54.00$\pm$3.50&311\\ \hline
03 May 03 & 62.498 &0.200 &27.54$\pm$0.66&96\\
 &  62.502&0.200 &34.03$\pm$0.70&393\\
 &  62.507&0.201 &43.40$\pm$2.20&260\\
 &  62.512&0.201 &42.20$\pm$5.07&68\\
 &62.542&0.201 &13.37$\pm$0.45&80\\
 &62.545 &0.201 &8.47$\pm$0.58&122\\
 &62.554&0.202 &5.06$\pm$0.17&356\\
 &62.562&0.202 &15.32$\pm$0.33&416\\ \hline
10 May 03 & 69.551 &0.357 &44.30$\pm$1.80&435\\
 &69.556&0.357 &52.20$\pm$2.00&446\\
 & 69.561&0.357&56.40$\pm$2.10&455\\ \hline
11 May 03 & 70.492&0.377&8.30$\pm$0.45&258\\
 & 70.506& 0.378&2.92$\pm$0.40&116\\
 & 70.519& 0.378&1.30$\pm$1.40&45\\  \hline
\hline
\end{tabular}
\end{table}

Additionally, the calibrated $V^2$ values need to be corrected for the 
influence 
of the nearby component $\delta$Vel\,B.
The diffraction on the sky (through an individual VLTI 0.4\,m siderostat) of 
the fiber fundamental mode, which defines the interferometric field of view,  
is equivalent to an Airy disk with a 1\farcs38 diameter.
At the time of the observations
Aa+Ab and B were separated by $\sim(1.0\pm0.3)\arcsec$.
Depending on atmospheric conditions,  the interferograms are, therefore, contaminated by a random
and time varying fraction of light, i.e. an incoherent 
signal, from star B.
The visibilities must, accordingly, be multiplied by a factor:
\begin{equation}
V_{\rm c\/} =V\times (1+I_{\rm B\/}/I_{\rm Aa+Ab\/}) = (1.05\pm0.05)\times V
\end{equation}
$I_{\rm B\/}$ and $I_{\rm Aa+Ab\/}$ are the intensities collected by the interferometer from $\delta$Vel\,B and $\delta$Vel\,(Aa+Ab).
$I_{B}/I_{Aa+Ab}$ lies between 0 (no light from B) and $10^{-\Delta m/2.5}=0.09$ (star B is completely 
in the field of view),
where $\Delta m\sim2.6$ equals the K-band magnitude difference between B and Aa+Ab.

\subsection{Comparison to a model}\label{model}

   \begin{figure*}
   \includegraphics[width=16cm]{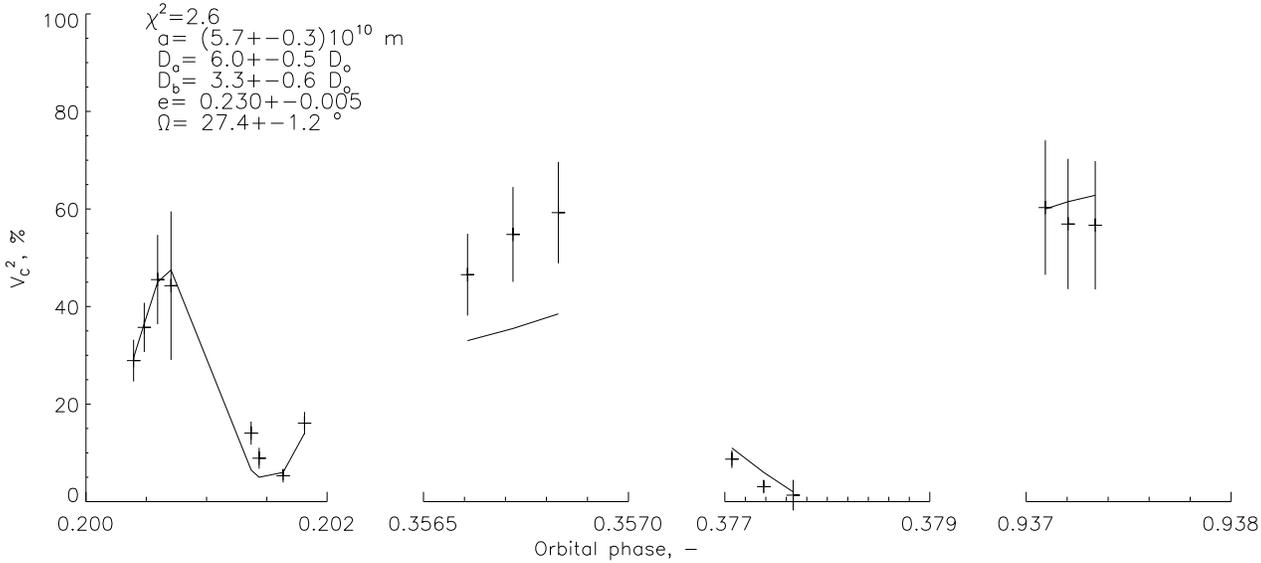}
      \caption{Corrected visibility values and standard deviation, confronted to 
      a model of two uniformly luminous, spherical stars. 
      The parameter values of the best fit (solid line) are indicated in the upper left corner.}
       \label{fit}
   \end{figure*}
   
   \begin{figure}
   \includegraphics[width=8cm]{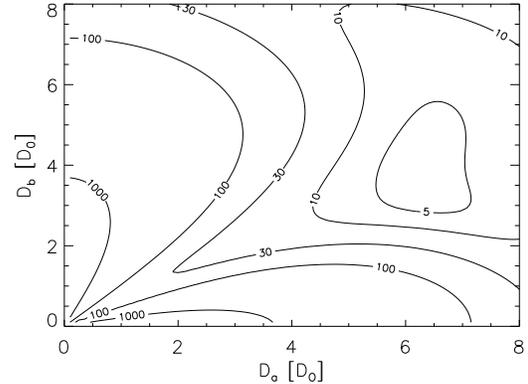}
      \caption{ $\chi^2$ as a function of the stellar diameters. The three other parameters of the model
      are set equal to: $a=(5.7\pm0.3)10^{10}$m, $e=0.230\pm0.05$, $\Omega=27.4\pm1.2^\circ$.}
       \label{chi2}
   \end{figure}

The 17 visibility measurements, ${V_c}^2$, were fitted to a model of
a binary system of two uniformly bright spherical stellar discs,
observed at K-band with a filter of finite bandwith.
Five parameters of the binary model (stellar diameters $D_a,
D_b$, position angle of the Ascending Node $\Omega$, semi-major axis $a$,
eccentricity $e$) were adjusted for optimum fit  to the observations.
The fitting procedure utilizes a non-linear least-squares algorithm 
(Markwardt 2005) that
follows the direction of steepest descent of $\chi^2$ 
in the parameter space,
$\chi^2$ being the reduced sum of squared deviations, i.e. the sum divided by the 13 degrees of freedom. To distinguish between local and absolute minima, the initial parameters were varied over the broad ranges of their potential values: 
The semi major axis, $a$, was considered between $5.4\,10^{10}$\,m and $8.0\,10^{10}$\,m, 
which corresponds to a total mass of Aa and Ab in the range $3 - 10\,M_\odot$. 
As specified in Sec.\,\ref{Photometry}, $e\in[0.23, 0.37]$. 
The stellar diameters were examined between 0.4 and 12.4\,mas. These limits 
refer respectively to the resolution limit of the interferometer and to 
the Roche lobe volume diameter $D_{L}$. The latter is approximated to better 
than 1$\%$ by $D_{L}/d\sim 12.4$\,mas (Eggleton\,\cite{Eggleton1983}).
If one of the stars were to have a diameter larger than $D_{L}$, the system 
would be an 
interacting binary and
the simple model of two spherical, uniformly bright stars would not apply.
The position angle of the Ascending Node $\Omega$, measured from North to
East, equals 0 if the projected
orbital plane and the North-South axes are aligned.
No previous measurement of $\Omega$ exists, and the angles $\Omega$ and 
$\Omega+\pi$ can not be distinguished through interferometric measurements, 
$\Omega$ is therefore considered between 0 and $\pi$.
Varying the surface brightness ratio $\phi$ over the
range specified in Eq.\,\ref{eq:phi} has virtually no effect on $\chi^2$, and $\phi$
is therefore fixed at 1.46.
Likewise, the period of the binary $\delta$\,Vel\,(Aa-Ab) is
fixed at $T=45.150$\,days. No apsidal motion of the eclipsing system 
has been noted since its discovery in 2000.
The orbital inclination has been fixed at $90^\circ$, 
although,  given the stellar diameters and separations deduced in Section\,\ref{Stellarparameters}, 
the actual inclination could lie between $87.5^\circ$ and $92.5^\circ$.

The best adjustment of the model to the 
measured visibilities and their 1-sigma statistical errors, is shown on Fig.\,\ref{fit}.
It corresponds to a reduced mean squared deviation
$\chi_0^2=2.6$, and is obtained for the following parameter values:
$ a = (5.7\pm0.3)\times10^{10} {\rm m\/}, 
e = 0.230\pm0.005, 
\Omega = (27.4\pm1.2)^{\circ}, 
D_a = (6.0\pm0.5)\,D_\odot, 
D_b = (3.3\pm0.6)\,D_\odot.$
The angle at primary eclipse is derived by the eccentricity as specified in Section\,\ref{Orbitalelements}: $\omega=-(20\pm3)^{\circ}$.
The parameter uncertainties equal the statistical errors, $\sigma$, scaled by the reduced mean deviation of the model to the measurements, i.e. $\chi_0\,\sigma$. 
The dependence of $\chi^2$ on the stellar diameters is illustrated on Fig.\,\ref{chi2}.

The three visibilities measured  on May 10, 2003 systematically deviate from the model fit (see Fig.\,\ref{fit}). 
There is no evident explanation for this deviation: 
the data were obtained under good atmospheric conditions and the calibrator was the same as on the other nights. 
If the three points are removed, the quality of the fit is improved, $\chi_0^2=1.4$, 
but within the uncertainties, the resulting parameter values are unchanged:
$ a = (5.4\pm0.5)\times10^{10} {\rm m\/}, 
e = 0.230\pm0.005, 
\Omega = (29.2\pm2.4)^{\circ}, 
D_a = (6.6\pm0.5)\,D_\odot, 
D_b = (3.2\pm0.5)\,D_\odot.$

It is apparent from the relatively high  $\chi_0^2$ that there are deviations in addition to the purely statistical errors. They might be due to an underestimate of the calibrator's size or might reflect some inaccuracies of the model of two uniformly bright, spherical stars. This is discussed in the subsequent section.

\section{Results and discussion}\label{Discussion}
\subsection{The close eclipsing binary $\delta$\,Vel\,(Aa-Ab)}

The computations could be slightly biased if the diameter of the calibrator star were substantially misestimated, or if HD\,63744 were a  -- still undiscovered -- binary system. 
On the other hand, HD\,63744 is part of the catalog of interferometric calibrator stars 
by Bord\'e et al.\,\cite{Borde}, with its diameter ($1.63\pm0.03$)\,mas specified to 
 a precision of $1.8\%$.
Furthermore, it has been studied simultaneously with other calibrator stars in VINCI observations by one of the authors (P.\, Kervella). In these investigations the visibilities of HD\,63744 equaled those expected for a single star of $1.63\pm0.03$\,mas diameter.
Thus, HD\,63744 appears to be a reliable calibrator.

A perhaps more relevant aspect are possible astrophysical complexities of $\delta$\,Vel\,(Aa+Ab)
that are disregarded in the model of two uniformly bright, spherical stars. In particular, the rotational velocities of Aa and/or Ab are found to be high, with values of $\sim$150 - 180\,km/s (Royer et al.\ \cite{Royer},
Hempel et al.\ \cite{Hempel}, Holweger et al.\ \cite{Holweger}), 
which indicates that the two stars need not be uniformly luminous nor circular.

Another possible over-simplification of our binary model is the constraint on the orbital inclination, $i$, being fixed at 90$^\circ$. 
Given the fitted semi-major axis and stellar diameters, we note that the eclipse durations ($0.51\pm0.05$\,days and $0.91\pm0.01$\,days) are shorter than they should be in the case of  $i=90^\circ$,
where the duration of the longer eclipse would have to exceed  $D_a\;T/(2\pi\,a)=1.06$\,days. 
We conclude that $i$ is $\sim 88^\circ$ or $\sim 92^\circ$, rather than $90^\circ$.
All observations were performed out of eclipse and, therefore, the visibility values are nearly unaffected by such a small variation in $i$. With substantially more visibility measurements and an increased number of fitted parameters, the issue on the precise orbital inclination might be addressed in more detail.

The most important and remarkable result of our analysis is that the stellar diameters of Aa and Ab are found to equal 
$6.0\pm0.5\,D_{\odot}$ and $3.3\pm0.6\,D_{\odot}$, respectively.
This exceeds significantly, by factors  $\sim$\,1.4 - 3, the values expected if Aa and Ab are main sequence stars. 
If both diameters are constrained to lie below $2.5D_{\odot}$, the
best fit corresponds to $\chi^2=16.7$, 
which is far beyond the present result and confirms that large diameters are required to account for the measured visibilities.

\subsection{On the physical association of $\delta$Vel~C and D}
Ever since the observations of Herschel (\cite{Herschel}),
$\delta$Vel has been taken to be a visual multiple star, 
$\delta$Vel~C and D being the  outer components of the system.
With m$_V$ of 11.0 mag and 13.5 mag 
respectively (Jeffers et al.\ \cite{Jeffers}), C and D would need to be of late 
spectral type,
certainly no earlier than M,
if they were at similar distance as $\delta$Vel~(Aa+Ab+B).
To our knowledge, the only existing spectra of C and D were recorded during
a survey of nearby M dwarfs (Hawley et al.\ \cite{Hawley}).
While the limited range and resolution of the spectra precluded ready
determination of the spectral types of C and D, they were nevertheless
estimated as $\sim$G8V and $\sim$K0V. Therefore, given their apparent
magnitudes, C and D must be at much further distances than
$\delta$Vel~(Aa+Ab+B).
We conclude, that $\delta$Vel~C and D are not physically associated.
Hence, $\delta$Vel ought to be classified as a triple stellar system
only.


\section{Summary}\label{conclusion}

Seventeen VINCI visibility measurements of $\delta$\,Vel\,(Aa+Ab) were fitted onto the 
model of two uniformly bright, spherical stars. The adjustment to the measurements does not provide individual diameters compatible with A-type main sequence stars. The two stars appear, thus, to be in a more advanced evolutionary stage. More data are, however, needed to ensure this result.
As the stellar evolution is fast during this period, more detailed knowledge of the system might also constrain the models more tightly. Precise photometric and spectroscopic observations of the eclipses should provide the separate intensities and chemical compositions of Aa and Ab and, hence, permit further inferences on the age and evolutionary state of $\delta$\,Vel.

\begin{acknowledgements}
We thank Rosanna Faraggiana for her extensive help, Sebastian Otero for providing the 
light curves $\delta$\,Vel Aa, Ab and Neil Reid for making available the spectra of $\delta$\,Vel C, D.
The manuscript was improved by helpful comments from the referee.
\end{acknowledgements}

\end{document}